\documentclass[
reprint,
amsmath,amssymb,
pra,
]{revtex4-2}
\usepackage{physics}
\usepackage{graphicx}

\begin{document}

\title{Robust Atom Optics for Bragg Atom Interferometry}
\author{Garrett Louie}
\author{Zilin Chen}
\author{Tejas Deshpande}
\author{Tim Kovachy}
\affiliation{%
 Department of Physics and Astronomy and Center for Fundamental Physics, Northwestern University, Evanston, IL 60208}%
 
\begin{abstract}
    Multi-photon Bragg diffraction is a powerful method for fast, coherent momentum transfer of atom waves. However, laser noise, Doppler detunings, and cloud expansion limit its efficiency in large momentum transfer (LMT) pulse sequences. We present simulation studies of robust Bragg pulses developed through numerical quantum optimal control. Optimized pulse performance under noise and cloud inhomogeneities is analyzed and compared to analogous Gaussian and adiabatic rapid passage (ARP) pulses in simulated LMT Mach-Zehnder interferometry sequences. The optimized pulses maintain robust population transfer and phase response over a broader range of noise, resulting in superior contrast in LMT sequences with thermal atom clouds and intensity inhomogeneities. Large optimized LMT sequences use lower pulse area than Gaussian pulses, making them less susceptible to spontaneous emission loss. The optimized sequences maintain over five times better contrast with tens of $\hbar k$ momentum separation and offers more improvement with greater LMT. Such pulses could allow operation of Bragg atom interferometers with unprecedented sensitivity, improved contrast, and hotter atom sources.
\end{abstract}

\maketitle

\section{Introduction}
Light pulse atom interferometry is a versatile and powerful platform for precision metrology and sensing, with applications ranging from fundamental physics \cite{Zhou2015_equivalence, rosi2017equivalence, Overstreet2018_equivalence, Asenbaum2017, Asenbaum2020_equivalence, fray2004atomic, Schlippert2014ep, Tarallo2014EquivSr, Barrett2016, kuhn2014bose, PhysRevA.88.043615, Hartwig2015, williams2016quantum, barrett2015correlative, overstreet2022observation, Bouchendira2011_finestructure, parker2018finestructure, morel2020determination, Biedermann2015_gravity, rosi2014precision, Arvanitaki2018_darkmatter, Graham2016_darkmatter,banerjee2022phenomenology, hamilton2015_darkenergy} to gravitational wave detection \cite{Dimopoulos2008_GW, graham2013new, Graham2016_GW, PhysRevD.93.021101, canuel2018exploring, hogan2011atomic, abou2020aedge, Badurina_2020, ZAIGA2020, abe2021matter} to mobile inertial sensors \cite{bidel2018absolute, bongs2019taking, Wu2019_mobile}. The most sensitive interferometers utilize large momentum transfer (LMT) atom optics over a long baseline to maximize the enclosed spacetime area. 

A proven technique for rapidly transferring many photon recoils is multi-photon Bragg diffraction \cite{Giltner1995Bragg, Martin1988Bragg, muller2008_24hk, Mazzoni2015, kovachy2015_halfmeter}, where a $2n$-photon transition in an optical standing wave couples momentum states separated by $2n\hbar k$. However, the population transfer and imprinted laser phase of Bragg atom optics are highly sensitive to the multiphoton detuning (e.g. from Doppler shifts) and to errors in the Rabi frequency (e.g. from laser instability or cloud expansion into regions of different laser intensity). Inhomogeneities in the atom cloud and the laser pulses therefore limit the achievable Bragg diffraction order and the number of pulses in a sequence. Spontaneous emission loss, which is mitigated by larger excited state detunings, further restrict the pulse area and maximum Rabi frequency for a given available laser power. 

One promising solution is to engineer more advanced Bragg pulses that are robust to such variations using quantum optimal control techniques. In atom interferometry, quantum control schemes including composite pulses \cite{Butts2013, dunning2014_composite, Berg2015_composite}, shaped pulses \cite{emsley1992optimization, Luo2016_shaped_pulses}, adiabatic rapid passage (ARP) \cite{Kotru2015}, and numerical optimal control \cite{saywell2018_QuOC_Mirror, saywell2020optimal, saywell_2020_biselective} have been applied to Raman transitions with alkalis, while Floquet pulse engineering has been applied to single-photon atom optics on the $^1{\rm S}_0 \to {}^3{\rm P}_1$ transition of Sr \cite{wilkasonFloquet}. Existing optimal control applications to Bragg diffraction have explored numerical optimization of single order pulses in combination with higher order ARP \cite{kovachy2012ARP, goerz2021quantum, Goerz2023}. However, higher order Bragg diffraction has received limited attention through numerical optimal control \footnote{While this manuscript was in preparation, related work on this topic came to our attention \cite{BiercukSPIE, anderson2023improving, saywell2023enhancing}, including the first experimental verification of robust control of Bragg transitions.}, and adiabatic rapid passage suffers from a poor phase response to amplitude and detuning errors, which can easily wash out the interference signal \cite{kovachy2012ARP}. 

Here, we report simulation studies of numerical optimal control's application to higher order Bragg diffraction. Phase and amplitude modulated pulses are developed to drive LMT mirrors and beamsplitters, transferring up to $20 \hbar k$ per pulse. The optimized pulses maintain both good population transfer and stable phase shifts over a broad range of errors, resulting in improved interferometer contrast with comparable or lesser pulse area to analogous Gaussian sequences and order of magnitude lower pulse area than ARP sequences.

Successful implementation of such pulses would enhance sensitivity of atom interferometers through increased contrast, momentum separation, and interrogation times, facilitating new achievements in both fundamental physics and sensing. Our results are presented as using the 461 nm transition in $^{88}{\rm Sr}$ ($f_r = \omega_r/2\pi = 10.7$ kHz) but are easily adaptable to other transitions.

\section{System and Optimization}
\subsection{Interferometry Model}
Bragg transitions are driven by two counter-propagating beams of frequencies $\omega_{\rm up}$ and $\omega_{\rm down}$ and relative phase $\phi = \phi_{\rm up}-\phi_{\rm down}$, each with detuning $\Delta$ from the optical excited state. After adiabatic elimination of the excited state assuming large detuning \cite{MeystreAtomOptics}, the rotating frame Hamiltonian can be written in the Hilbert space of momentum eigenstates $\ket{2l\hbar k}$ for integer $l$. In a frame where the atom is stationary and the beams have frequency difference $\omega_0 \equiv \omega_{\rm up} - \omega_{\rm down}$, the matrix elements are \cite{kovachy2010lattice, Muller2008Bragg} 
\begin{equation}
\begin{split}
    H_{lm}/\hbar &= \left[4l^2\omega_r - l(\omega_0+\alpha)\right]\delta_{l,m} \\
    &+ \frac{1}{2}(1+\eta)\Omega(t) \left[e^{-i\phi(t)}\delta_{l,m-1} + {\rm H.c.}\right]
\end{split}
\label{eqn:Hamiltonian}
\end{equation}
where $\omega_r \equiv \frac{\hbar k^2}{2m}$ is the recoil frequency, $\Omega(t) \equiv \frac{\Omega_{\rm up}\Omega_{\rm down}}{2\Delta}$ is the two-photon effective Rabi frequency, $\alpha \equiv 2k\var p/m$ is a detuning caused by momentum offset $\var p$, and $\eta$ is the fractional error in $\Omega$. 

Since $\omega_0$ is not an optimization variable, the Hamiltonian used for optimization of an $n$-th order pulse is written in a frame falling relative to the atoms at such a velocity that the desired transition is on resonance when $\omega_0=0$. We include the $n+1$ discrete momentum states padded on each end by three levels (i.e., $\ket{-6\hbar k},\ket{-4\hbar k},\ldots, \ket{(2n+6)\hbar k}$ with initial state $\ket{0\hbar k}$). For sequence simulations, we include all the momentum states within the range of the sequence, padded on each end by at least five levels. In this treatment, undesired interferometer paths are discarded for simplicity; the output ports are assumed to be spatially separated enough that these paths should not significantly affect the signal. Furthermore, we note that the optimized pulses would only reduce the effect of stray paths. 

\begin{figure}
    \centering
    \includegraphics[width=\linewidth]{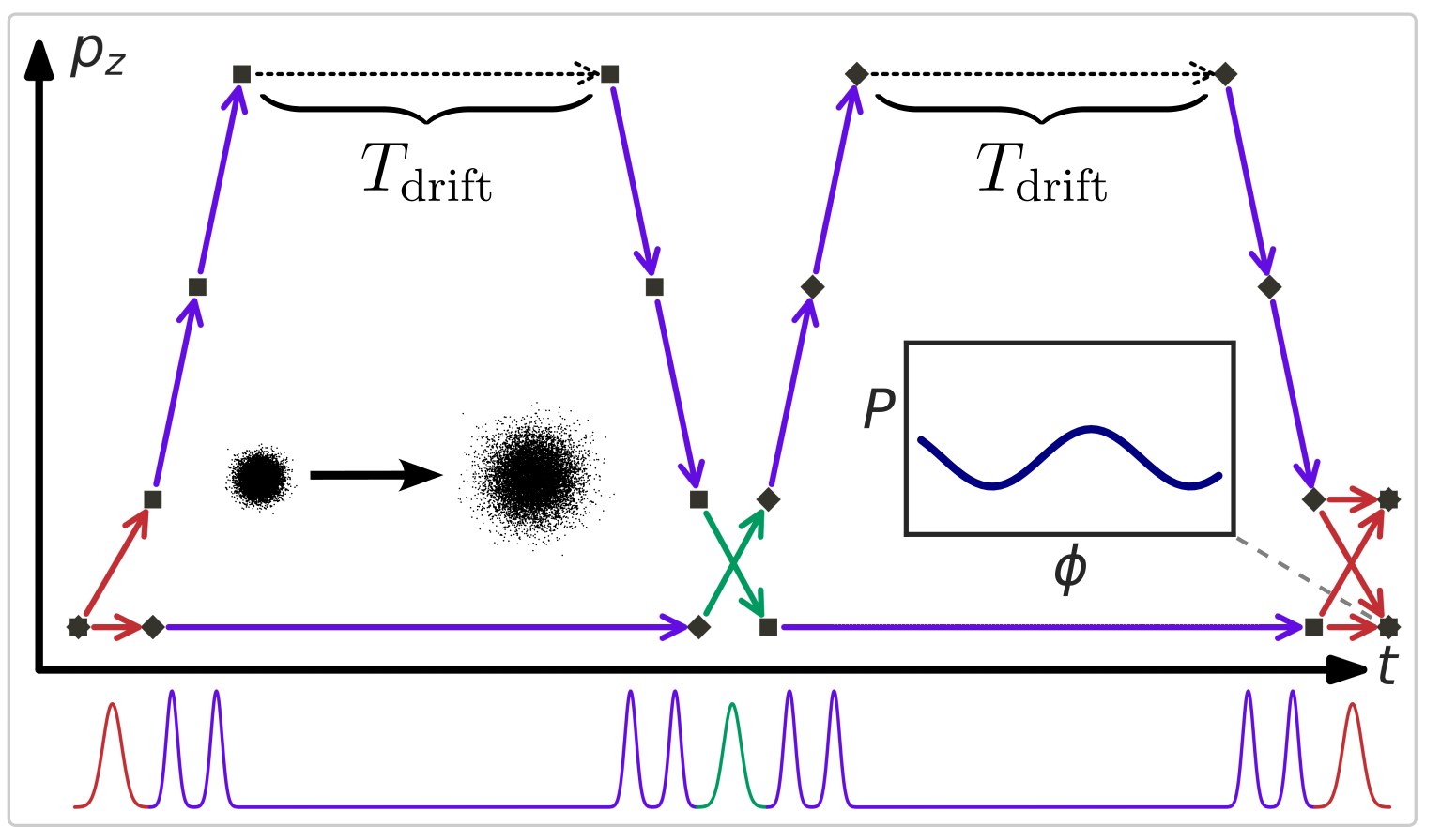}
    \caption{Momentum-time diagram of the two arms (squares and diamonds) in an LMT Mach-Zehnder interferometry sequence, showing timing of beamsplitter (red), mirror (green), and augment (purple) pulses. At peak momentum separation, the arms spatially separate in free fall for a period $T_{\rm drift}$, during which the cloud expands. The contrast is extracted by scanning the phase of the final beamsplitter.}
    \label{fig:LMTsequence}
\end{figure}

We simulate Mach-Zehnder LMT sequences using three pulse types: a beamsplitter, a mirror, and an augment, of diffraction orders $n_{BM}$ and $n_A$ respectively. The beamsplitter pulse splits the momentum into two arms $2n_{BM}\hbar k$ apart, and a sequence of $m$ augment pulses increases the separation by $2mn_A\hbar k$. After a period of drift to allow the arms to spatially separate, the augment sequence is reversed and the mirror pulse swaps the momenta of the two arms. The augment-drift-augment sequence is repeated on the second arm, after which a final beamsplitter pulse recombines the arms (Figure \ref{fig:LMTsequence}). The affect of pulses on the off-resonant arm is included in the simulations. 

To extract fringe contrast, population in one output port as a function of the final beamsplitter phase is averaged over the noise distribution and fit to the form
\begin{align}
    P(\phi) = P_0 + \frac{C}{2}\cos(n_{BM}\phi+\var\phi)
\end{align}
where $C$ is the contrast and $\var\phi$ is a variable phase shift from averaging interferometer errors. 

The spontaneous emission loss is $1-e^{-\frac{\Gamma\mathcal{A}}{\Delta}}$, where $\mathcal{A}\equiv \int\Omega(t){\rm d}t$ is the two-photon effective pulse area \cite{metcalf_book, LesHouches_notes}. To compare susceptibility to spontaneous emission, we note the ratios in total pulse area between each sequence in our simulations. 

\subsection{Optimal Control Method}
We determine the optimized pulse waveforms with ensemble optimization \cite{Goerz2014Ensemble} using the optimizer in Q-CTRL's Boulder Opal package \cite{ball2021software}: robust control is achieved by averaging fidelity over a batch of perturbed Hamiltonians that sample an experimentally relevant distribution of noise values. We calculate the unitary evolution $\hat{U}^{(i)}$ for $N_{\rm batch}$ Hamiltonians of the form of Equation \ref{eqn:Hamiltonian}, each with values of $\alpha$ and $\eta$ randomly sampled from Gaussian distributions centered at zero whose predefined width determines the scope of desired robustness. For each, we calculate an infidelity $\mathcal{I}^{(i)}$. For atom interferometry, this metric should be sensitive to both the population and phase of the target state(s), since inhomogeneous phase shifts may cause contrast loss. For augment pulses, we use
\begin{equation}
    \mathcal{I}^{(i)}_{\rm state} = 1 + \Im\left[\mel{2n\hbar k}{\hat{U}^{(i)}}{0\hbar k}\right]
    \label{eqn:stateinfidelity}
\end{equation}
since only transfer to the desired state is important. For beamsplitters and mirrors, we use
\begin{equation}
    \mathcal{I}^{(i)}_{\rm gate}=1-\left|\frac{\operatorname{Tr}\left(U_{\text {target }}^{\dagger} U^{(i)}\right)}{\operatorname{Tr}\left(U_{\text {target }}^{\dagger} U_{\text {target }}\right)}\right|^{2}
    \label{eqn:gateinfidelity}
\end{equation}
since these pulses perform general rotations on populations in both the initial and target states. The total cost function is the average over the batch of infidelities and thus is minimized when fidelity remains good in the presence of noise. 

The optimization variables correspond to time segments of the control waveforms $\Omega(t)$ and $\phi(t)$. However, since leakage to neighboring diffraction orders is the primary cause of atom loss, we have found it beneficial to limit the spectral width of the pulses. Therefore, the series of constant segments defined by the optimization variables is smoothed with a sinc filter, imposing a frequency domain cutoff frequency of $4n\omega_r$ for an $n$-th order pulse. This procedure is expedited by the convolution tools in Boulder Opal \cite{ball2021software}. 

The pulse duration, sampling parameters, maximum Rabi frequency, and batch size were set manually before optimization. The results presented in this paper used a batch size $N_{\rm batch} = 256$ and 128 time segments for each of $\Omega(t)$ and $\phi(t)$, limited by available computing resources through Boulder Opal. Values of amplitude and momentum errors were sampled from Gaussian distributions of width 0.05-0.1 and 0.15-0.25 $kv_r$ respectively. For the augment pulse in Figure \ref{fig:waveform}, the batch of infidelities had an additional weighted average over three detuning values at $0$ and $\pm 0.15kv_r$, which effectively broadened the sampling distribution. Duration and maximum Rabi frequency were chosen to make pulse area comparable to Gaussian pulses and assuming laser power suitable for relevant atom optics laser systems. That is, the Rabi frequencies are achievable with under 20\% spontaneous emission loss (Figure \ref{fig:FinalFigure}) in Sr using millimeter to centimeter scale beam waists and Watts of laser power. 

\section{Results}

\begin{figure}
    \centering
    \includegraphics[width=\linewidth]{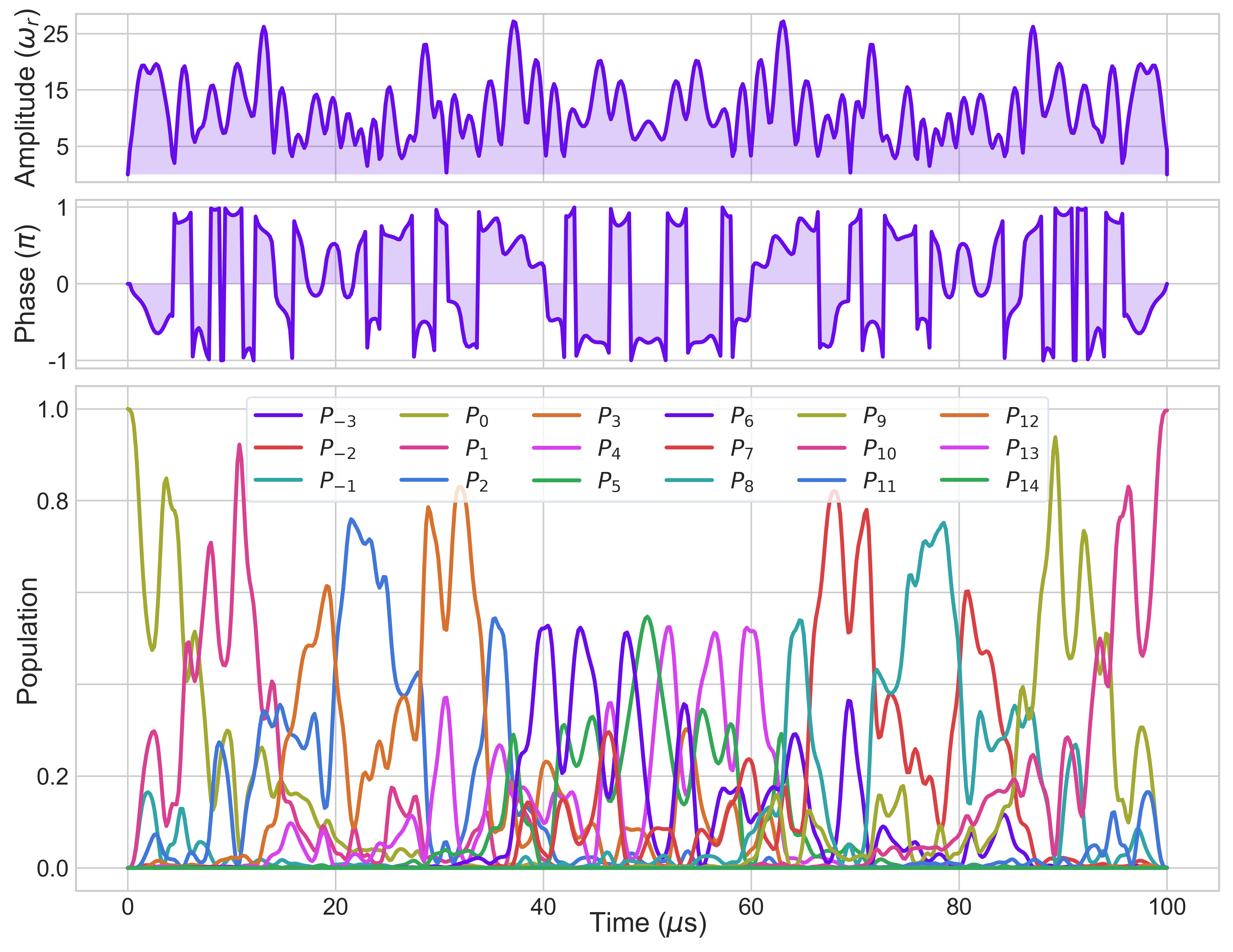}
    \caption{Optimized amplitude and phase drives (top), with evolution of momentum state populations (bottom) for an $n=10$ augment pulse.}
    \label{fig:waveform}
\end{figure}

We have developed optimized mirror pulses for Bragg diffraction order up to $n=10$, and beamsplitter pulses up to $n=8$, similarly limited by available computing resources through Boulder Opal. The phase and amplitude waveforms along with state evolution for a tenth order augment pulse are shown in Figure \ref{fig:waveform}. Since the augment pulses are intended to be repeated in longer interferometry sequences, the duration was limited to $100 \mu s$ ($6.7/\omega_r$). Beamsplitter and mirror pulses are $100 \mu$s and $80 \mu$s long with lower amplitudes of $10.2\omega_r$ and $10.9\omega_r$ respectively, which were appropriate for the lower order. 

\begin{figure}
    \centering
    \includegraphics[width=\linewidth]{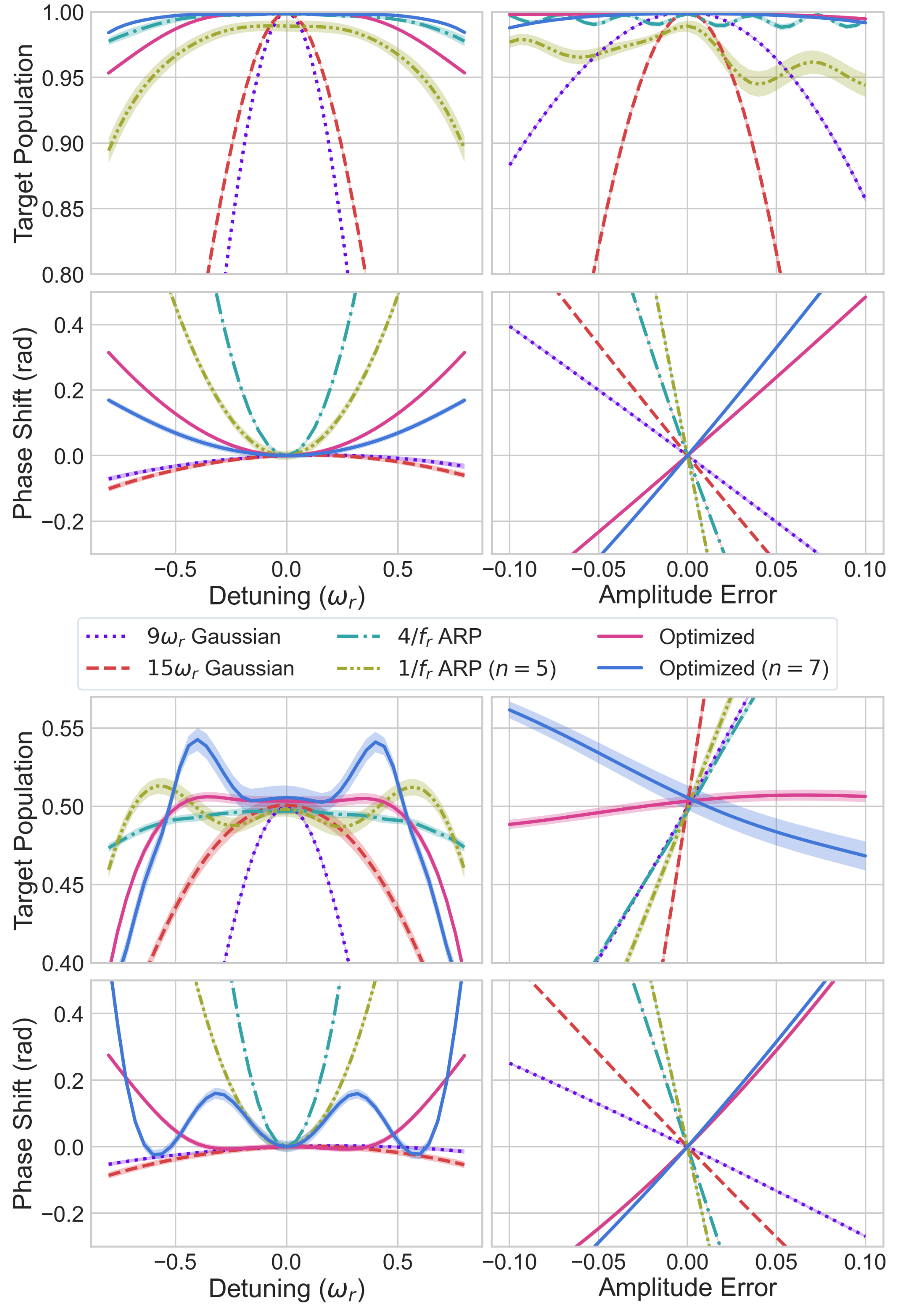}
    \caption{Response to static momentum (left) and amplitude (right) errors for $n=3$ mirror (top) and beamsplitter (bottom) pulses, including spectral noise. Shaded bands show statistical standard deviation over an ensemble of 50 stochastic phase and amplitude noise trajectories sampled from a white spectrum from 100 Hz to 1 MHz, normalized to an RMS of 5 mrad and 0.5\% respectively. A set of optimized $n=7$ pulses is also included to show that optimizing higher order beamsplitters is possible with these methods.}
    \label{fig:1Dscans}
\end{figure}

\begin{figure*}
    \centering
    \includegraphics[width=\textwidth]{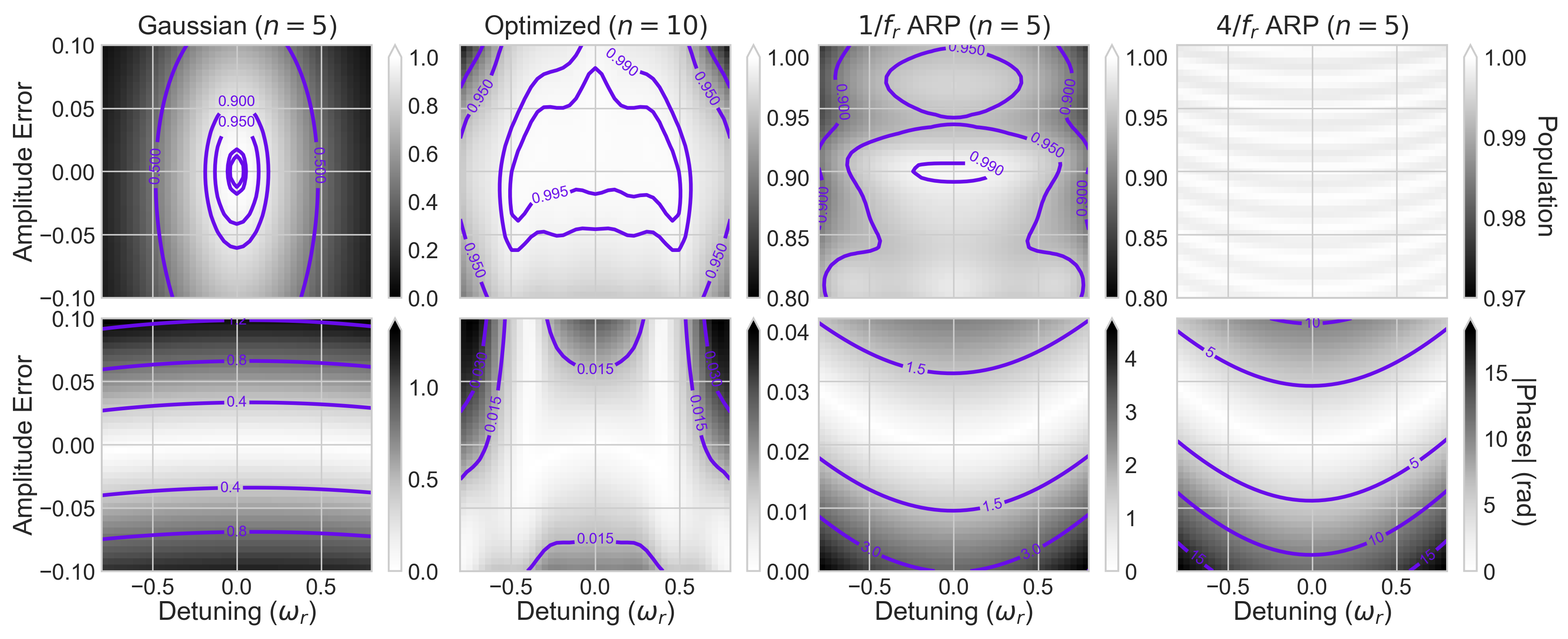}
    \caption{Response of augment pulses to simultaneous amplitude and detuning errors, showing population transfer (top) and magnitude of the phase deviation from the noiseless value (bottom). All pulses have a maximum two-photon Rabi frequency of $27.2 \omega_r$, except the $1/f_r$ ARP, which has its maximum transfer efficiency at $29.2\omega_r$.}
    \label{fig:2Dscans}
\end{figure*}

To visualize the pulse response to noise, we compute the statevector evolution under the relevant range of noise values. Figure \ref{fig:1Dscans} shows the population and phase response of $n=3$ mirror and beamsplitter pulses optimized using Equation \ref{eqn:gateinfidelity}, compared to several analogous Gaussian and ARP pulses. Simple square pulses are not applicable as their broad spectrum mandates much smaller Rabi frequencies and longer interaction times \cite{Muller2008Bragg}. Although not shown in the figures, we found that pulses optimized using the alternative infidelity metric $\mathcal{I}_{\rm state} + \mathcal{I}_{\rm gate}$ had a better phase response than $\mathcal{I}_{\rm gate}$ alone. That said, this did not produce a significant improvement in contrast over the range of errors considered in the sequence simulations. 

To include the effect of incoherent laser noise inherent to physical systems, the results are averaged over an ensemble of 50 stochastic phase and amplitude noise trajectories sampled from a white spectrum between 100 Hz and 1 MHz, which are normalized to an RMS of 5 mrad and 0.5\% respectively. Below this range noise is effectively static, while higher frequencies have little effect. A level of noise at or below this magnitude can be achieved by the laser systems used in atom interferometry \cite{Muller:06_lasers}. We have verified, through analysis of stochastic noise sampling and through computation of filter functions in Boulder Opal \cite{ball2021software}, that despite being optimized against static amplitude and momentum errors, our pulses are also no more susceptible to spectral amplitude and phase noise than Gaussian pulses and significantly less susceptible than ARP pulses.

The ARP pulses have an envelope of the form $\Omega(t) = \tanh(8t/T)\tanh[8(1-t/T)]$ and a linear frequency sweep from $-2\omega_r$ to $2\omega_r$ for duration $T$, as was demonstrated experimentally in \cite{kovachy2012ARP}. Longer ARP duration and a broader frequency sweep lead to better robustness in population transfer, at the cost of greater phase shifts, pulse area, and susceptibility to laser noise. We show the noise response for three ARP pulses with different durations: $8/f_r$ and $4/f_r$ with $\Omega_{\rm max}=10\omega_r$; and $1/f_r$ with $\Omega_{\rm max}=29.2\omega_r$, the $n=5$ mirror pulse used in \cite{kovachy2012ARP} (a $1/f_r$ pulse is too short for $n=3$ due to the closer energy level spacing). Beam splitters have appropriate lower amplitudes with the same durations. 

Two pairs of Gaussian pulses are also shown. The lower amplitude pulses have a width $\sigma = 1/\omega_r$ with amplitude $9.2 \omega_r$ ($5.6\omega_r)$ chosen for $\pi$ ($\pi/2$) rotations. The higher amplitude pulses use an amplitude of $15\omega_r$ for both mirror and beamsplitter, with different durations chosen appropriately. Gaussian pulses with higher Rabi frequencies have better robustness to detuning errors but are more susceptible to amplitude errors. They may also have less pulse area since the coupling strength is a nonlinear function of amplitude and diffraction order \cite{Muller2008Bragg}. However, the pulse duration must be long enough that population is not lost to other diffraction orders. 

Lower power $n=3$ pulses are shown to provide examples at lower amplitudes than the following augment pulse comparison. For clarity, only the target state phase and population of the beamsplitters is shown, though we note that the initial state population varies opposite to that of the target state, and the noise response is similar when the initial and target states are reversed. Since the beamsplitter amplitudes lie at the linear points of Rabi oscillations, the population balance is inherently more susceptible to amplitude errors, so the optimized pulse offers a substantial improvement. We find that amplitude based phase shifts in mirrors and beamsplitters are generally common to the two arms\textemdash that is, $\mathcal{I}_{\rm gate}$ is unaffected\textemdash and do not cause a loss of contrast. However, in pulses which address only a single arm, such as augments, these phase shifts may have a significant effect. 

We have found that optimizing using Equation \ref{eqn:stateinfidelity} gives better results for augment pulses than the more stringent Equation \ref{eqn:gateinfidelity}, especially in the phase response. Figure \ref{fig:2Dscans} shows the population and the magnitude of the phase deviation as a function of simultaneous amplitude and detuning errors for higher order augment pulses of each type, with amplitudes chosen to match the optimized pulse ($27.2\omega_r$). Again, the ARP pulses maintain excellent population transfer across both noise channels, but the phase is volatile, especially for the longer $4/f_r$ pulse. The Gaussian pulse has a better phase response but poor population transfer under errors. The optimized pulse maintains good transfer and the best phase response across the range of noise. 

\begin{figure}
    \centering
    \includegraphics[width=\linewidth]{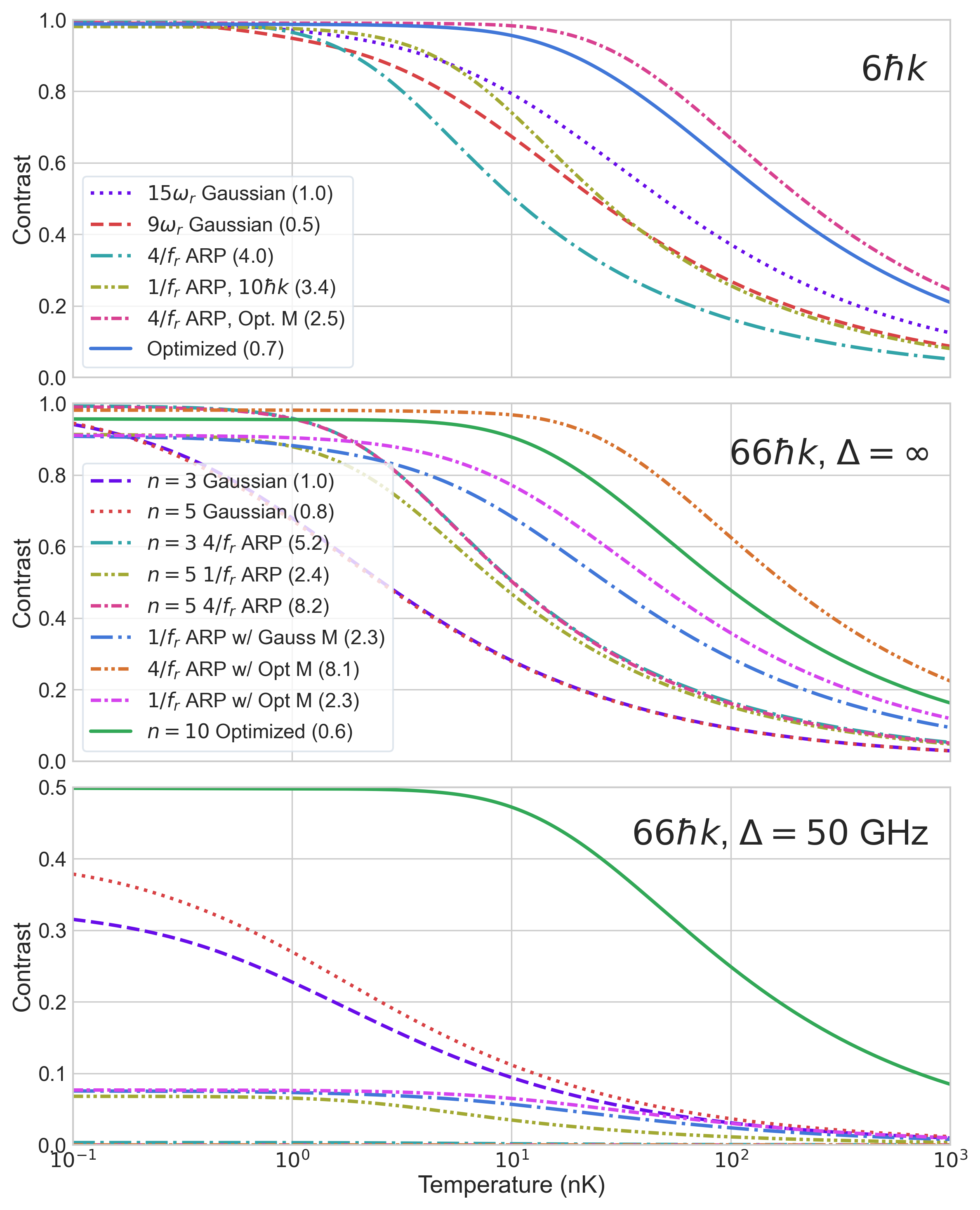}
    \caption{Contrast vs temperature vs LMT. (Top) Simple three pulse $6\hbar k$ beamsplitter-mirror-beamsplitter sequence using different pulse types. (Middle) $66\hbar k$ sequence with different augments. Gaussian sequences use the $15\omega_r$ mirror and beamsplitters, while ARP sequences use the $4/f_r$ mirror and beamsplitters unless otherwise indicated. (Bottom) Same as middle plot, but includes spontaneous emission loss with an excited state detuning of 50 GHz. Ratios of total pulse area to optimized sequence are indicated in legend parentheses. The sequence using the $1/f_r$ augment has a lower peak contrast because its noiseless transfer efficiency is only 99.3\%.}
    \label{fig:ContrastDetuning}
\end{figure}

To compute the effect of thermal Doppler detuning on LMT interferometer contrast, the sequence is simulated under a range of detunings and the output port is averaged over the 1D Maxwell-Boltzmann as a function of temperature. For the ARP sequence, an alternative using an optimized mirror is shown; since the mirror phase does not cancel in the sequence, the ARP mirrors have limited effectiveness. The optimized pulse sequence maintains superior contrast to Gaussian pulses, and in the longer sequence has a lesser pulse area. ARP beamsplitters and augments\textemdash for which detuning-based phase shifts do cancel\textemdash have slightly better transfer than the optimized pulse. However, the ARP sequences have significantly greater pulse area, as indicated by the legend values in parentheses, meaning a corresponding increase in spontaneous emission loss or required laser power. The bottom plot includes spontaneous emission loss with an excited state detuning of 50 GHz. In this regime, the optimized pulse maintains over five times better contrast at temperatures above 10 nK. The disparity would increase at greater LMT: losses are compounded with additional pulses, and the total pulse area of the optimized sequence decreases relative to the other schemes.

\begin{figure}
    \centering
    \includegraphics[width=\linewidth]{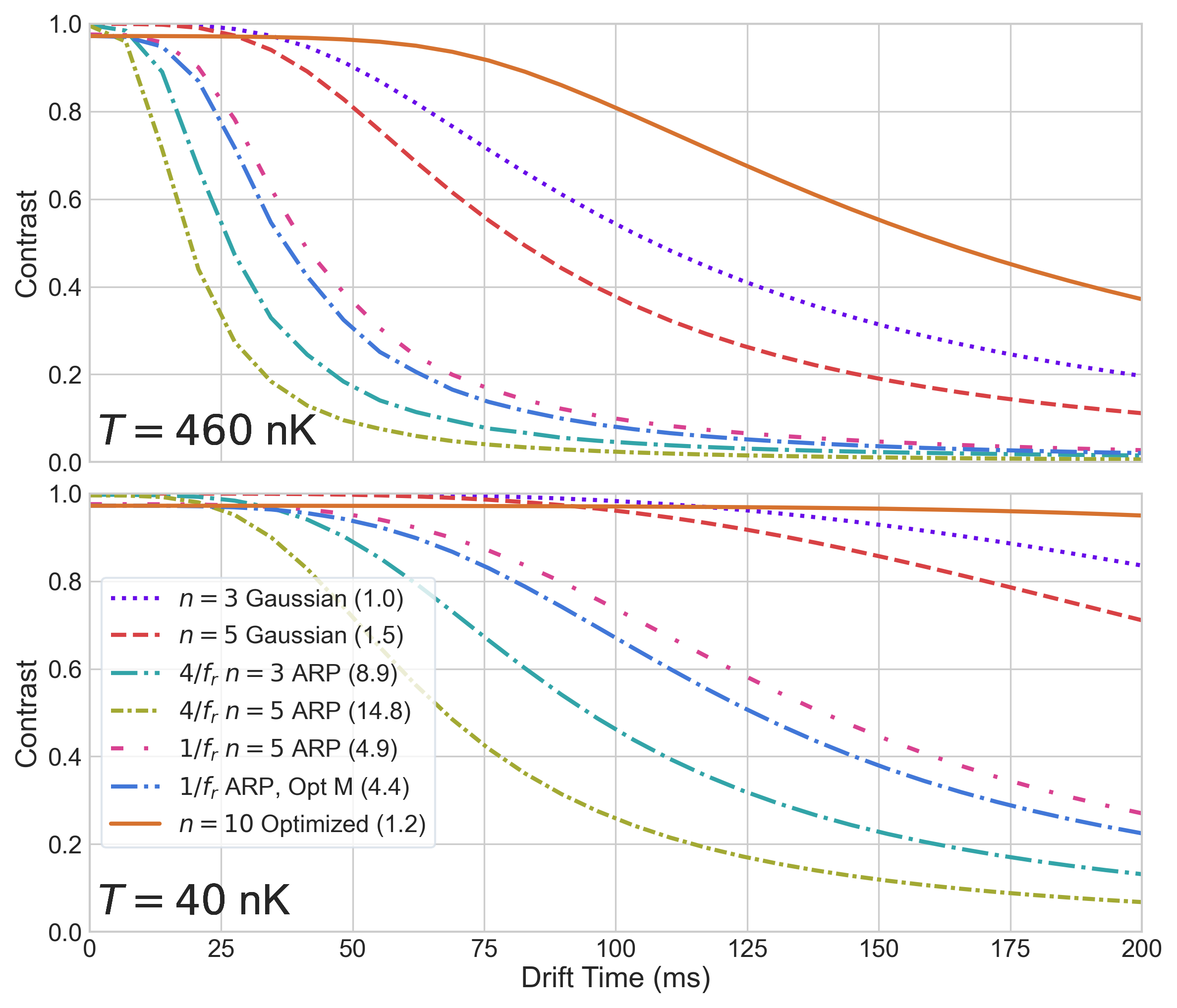}
    \caption{Contrast decay in sequences due to cloud expansion (ignoring Doppler shifts) with a 5 mm beam waist. Sequences using $n=3$ augments are $24\hbar k$, the rest are $26 \hbar k$. Gaussian sequences use the $9.2\omega_r$ ($5.6\omega_r$) mirror (beamsplitters), which are most robust to amplitude errors. In this case, using the optimized mirror in the ARP sequence does not help, since the mirror phase shift is common to the arms, and contrast loss instead stems from diffraction pahse of the augment pulses.}
    \label{fig:CloudExpansion}
\end{figure}

To illustrate the effect of amplitude errors, we consider the experimentally relevant case of cloud expansion in an LMT sequence with variable interrogation time. Figure \ref{fig:CloudExpansion} shows the contrast decay in a sequence with $18$-$20\hbar k$ augmentation for a two temperatures as a function of $T_{\rm drift}$ with a beam waist of 5 mm, ignoring detuning variation (this is not realistic, but it is informative to consider the effect of cloud expansion alone). The contrast is averaged over a 2D Maxwell-Boltzmann distribution of transverse velocities for each temperature. 460 nK is the recoil temperature of Sr, achievable through narrow line Doppler cooling \cite{stellmerthesis}. In contrast to detuning errors, the ARP sequences show the most rapid contrast decay due to intensity dependent phase shifts. Loss in the Gaussian sequence is again primarily due to transfer efficiency. The robust optimized sequences maintain contrast even with significant expansion.  We note that applying augment pulses symmetrically to both arms can suppress the phase shifts to some degree; this simulation highlights that significant contrast loss can arise from just a couple asymmetric augment pulses.

\begin{figure}
    \centering
    \includegraphics[width=\linewidth]{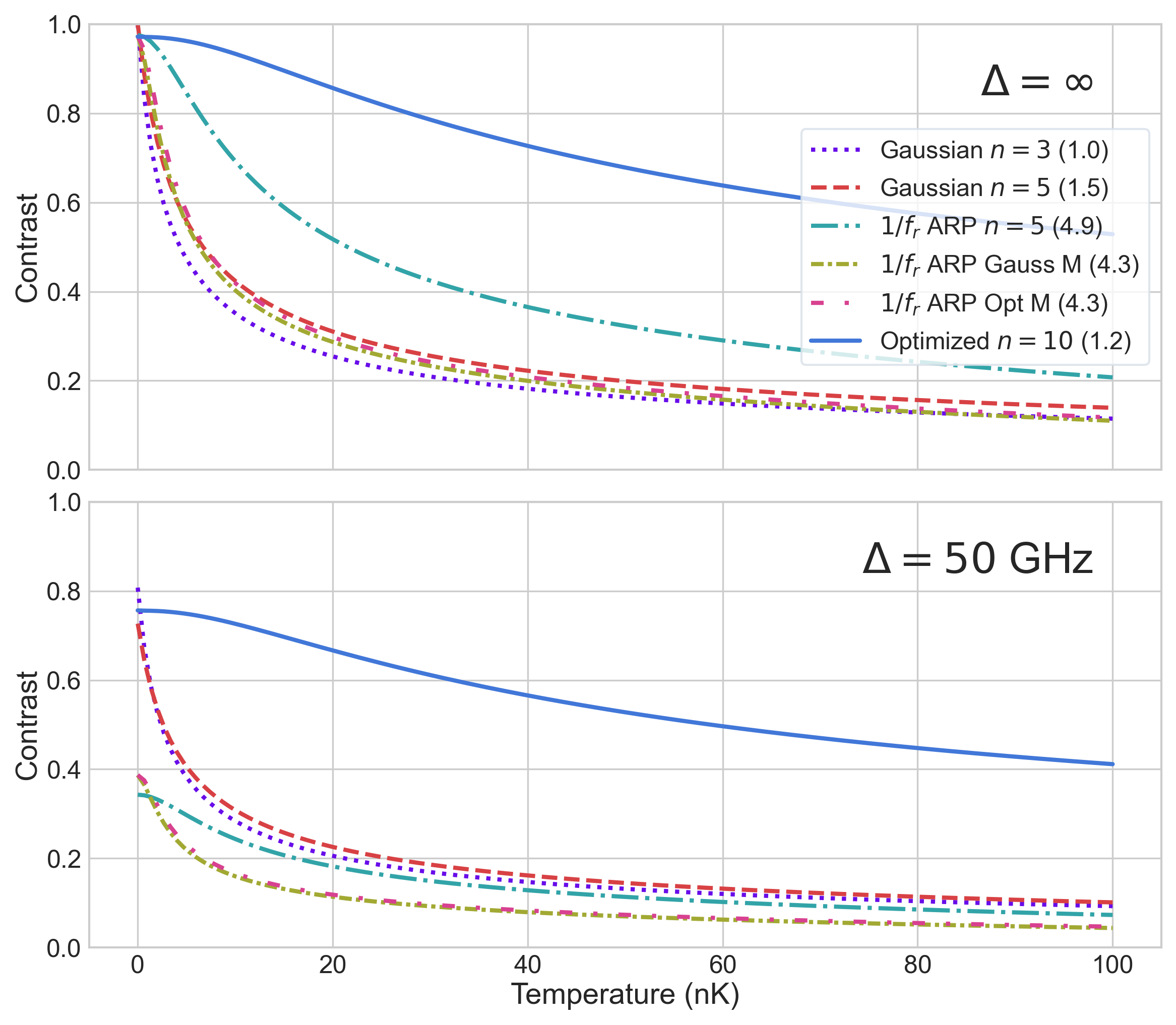}
    \caption{Contrast vs temperature in $24$-$26 \hbar k$ sequences with both cloud expansion and Doppler detunings for $T_{\rm drift}=50$ ms. Bottom plot is the same as top, but includes spontaneous emission loss using excited state detuning $\Delta=50$ GHz.}
    \label{fig:FinalFigure}
\end{figure}

Finally, we consider the combined effect of thermal detuning errors and cloud expansion. Figure \ref{fig:FinalFigure} shows the contrast decay in the same sequences as Figure \ref{fig:CloudExpansion}, assuming a beam waist of 5 mm and $T_{\rm drift}=100$ ms. The lower plot includes spontaneous emission loss, assuming an excited state detuning of 50 GHz; for this detuning and beam waist, a two-photon Rabi frequency of $28.5 \omega_r$ can be achieved in Sr with 1W of laser power in each beam. In this simulation, the optimized pulse maintains several times better contrast than all other schemes.

\section{Discussion}

We have explored the application of numerical optimal control to Bragg transitions in LMT atom interferometers. Pulses were engineered to be robust to variations in Rabi frequency and detuning, which we have shown to increase contrast in interferometry sequences. Their modest pulse area allows large LMT sequences with lower spontaneous emission losses, despite the pulse length increase generally required for optimal control. We have also examined the limitations of adiabatic rapid passage pulses, a quantum control method previously applied to Bragg diffraction.

The robustness to detuning errors allows use of hotter atom clouds, increasing signal and potentially shot rate. It is also useful for dual isotope interferometers where slight mass imbalances can cause slightly different velocity classes \cite{abe2021matter, fray2004atomic, Asenbaum2020_equivalence, Schlippert2014ep, Tarallo2014EquivSr, Barrett2016, kuhn2014bose, PhysRevA.88.043615, Hartwig2015, williams2016quantum}. Meanwhile, the robustness to pulse amplitude errors provides tolerance to laser noise and intensity variations across the beam, for instance due to cloud expansion and wavefront perturbations. Combined with the lower pulse area, this allows the use of narrower beam waists and smaller excited state detunings, reducing demands on laser power, which is a common limiting factor in scaling multi-photon Raman and Bragg pulse sequences 

We found that for a given maximum Rabi frequency, optimized pulses were able to provide robust and efficient transfer at a higher order than Gaussian pulses of comparable pulse area and duration. For example, compared to our $n=10$ augment pulse, at the same peak Rabi frequency ($27.2\omega_r$), an $n=5$ Gaussian pulse has around 70\% of the pulse area, but an $n=6$ Gaussian would require nearly 600\% the pulse area; the $n=3$ Gaussians with $9.2\omega_r$ and $15\omega_r$ amplitudes have 30\% and 50\% the pulse area respectively. A shorter optimized pulse could have been made which was still significantly more robust than the Gaussian, albeit to a lesser degree. As a result, tailored optimal control pulses would enable lower power laser systems to drive greater momentum transfer without any more spontaneous emission losses. 

Applying well-timed augment pulses symmetrically to both arms can suppress some of the amplitude based phase shifts. Nevertheless, pulses with a stable phase response allow more flexibility in interferometer geometry. For example, a single-arm momentum kick is necessary for a photon recoil measurement in a Ramsey-Bord\'e interferometer \cite{muller2006recoil, muller2008_24hk}. Moreover, factors such as beam divergence, pulse-to-pulse intensity fluctuations, and perturbations to the beam mode can easily break the symmetry, making phase-stable pulses easier to implement. 

The phase shift and hence the sensitivity of a Mach-Zehnder gravimeter scales as $\phi\sim kT^2$ for momentum separation $k$ and interrogation time $T$. For the most precise measurements\textemdash e.g., in tests of fundamental physics\textemdash it is therefore desirable to maximize both of these parameters. The next generation of long baseline atom interferometers seek momentum separations and interrogation times on the orders of $1000 \hbar k$ and 10 seconds respectively \cite{abe2021matter}. The tolerance to accumulated amplitude errors and the high diffraction order of the robust optimized pulses would allow significant increases in both momentum transfer and interrogation time, facilitating unprecedented sensitivity in state of the art atomic fountains. 

For atom interferometry applications in dynamic environments such as inertial sensing, short sequences may be desirable. Short, high power optimized pulses, particularly higher order beamsplitters, may provide quick contrast and sensitivity enhancements without compromising repetition rate. Alongside the robustness to laser fluctuations and the inherent insensitivity of Bragg transitions to the relative AC Stark shifts of the resonance between hyperfine levels that would appear in Raman transitions, this makes optimized Bragg pulses suitable for field sensing applications \cite{bidel2018absolute, bongs2019taking, BiercukSPIE, anderson2023improving}.

Other methods of optimization could also be investigated, including more complex cost functions and alternative methods of spectral filtering\textemdash for instance using a Fourier basis for optimization instead of time segments, or using smooth interpolation instead of convolution. We chose convolution primarily due to its convenience in the Boulder Opal package. 

In this work, we optimized individual pulses to perform under a broad range of noise and be ``modular" enough to combine into a variety of pulse sequences. Furthermore, our limits on the Rabi frequency were arbitrarily consistent with typical atom optics laser systems, assuming Watts of laser power and interrogation times of seconds. However, optimizing for a specific temperature, sequence, and/or laser system\textemdash e.g., with ensembles sampled from experimentally characterized noise distributions\textemdash would likely result in better results for a particular system. For instance, sequences with fewer pulses would favor robustness over peak efficiency. As an alternative approach, optimization of the entire interferometry sequence with contrast and sensitivity in the cost function, rather than optimizing single pulses, may also prove more effective, at the cost of generality. 

Future work will seek to experimentally demonstrate our results. These pulses require phase and amplitude modulation bandwidths of hundreds of kHz, readily achievable by standard acousto-optic modulators (AOMs). Ultimately, they may serve as a starting point for closed loop optimization \cite{Feng2018closedloop}, using experimental measurements to further adjust control waveforms. Successful closed loop protocols would enable more tailored control waveforms and on-demand calibration.

\begin{acknowledgements}
We thank Q-CTRL and Jens Koch for valuable discussions. This material is based upon work supported by the U.S. Department of Energy, Office of Science, National Quantum Information Science Research Centers, Superconducting Quantum Materials and Systems Center (SQMS) under contract number DE-AC02-07CH11359. 
\end{acknowledgements}

\bibliographystyle{apsrev4-2}
\bibliography{refs.bib}

\end{document}